\documentclass[prl,groupedaddress,reprint,superscriptaddress,floats,footinbib]{revtex4-1}
\usepackage{graphicx,gensymb}
\usepackage{amsmath,amssymb,amsfonts,mathtools}
\usepackage{hyperref}
\usepackage{slashed,wasysym}
\usepackage[usenames,dvipsnames]{xcolor}
\usepackage[normalem]{ulem}

\newcommand{\MET}{\slashed{E}_T}
\def\met{{\mbox{$E\kern-0.57em\raise0.19ex\hbox{/}_{T}$}}}
\newcommand{\mDM}{m_{\rm{DM}}}
\newcommand{\mA}{M_{A}}
\newcommand{\GamA}{\Gamma_{A}}

\newcommand{\gDM}{g_{\rm{DM}}}
\newcommand{\gSM}{g_{\rm{SM}}}

\newcommand{\AT}{\alpha_T}
\newcommand{\MT}{M_{T2}}
\newcommand{\Razor}{\textsc{Razor}}
\newcommand{\sigv}{\langle \sigma v \rangle}

\begin{document}

\title{Constraining Dark Matter Interactions with Pseudoscalar and Scalar Mediators\\ Using Collider Searches for Multi-jets plus Missing Transverse Energy}

\author{Oliver~Buchmueller}
\email[]{oliver.buchmueller@cern.ch}
\affiliation{High Energy Physics Group, Blackett Laboratory, Imperial College, Prince Consort Road, London, SW7 2AZ, United Kingdom}
\author{Sarah~A.~Malik}
\email[]{smalik@cern.ch}
\affiliation{High Energy Physics Group, Blackett Laboratory, Imperial College, Prince Consort Road, London, SW7 2AZ, United Kingdom}
\author{Christopher~McCabe}
\email[]{c.mccabe@uva.nl}
\affiliation{GRAPPA, University of Amsterdam, Science Park 904, 1098 XH Amsterdam, Netherlands}
\author{Bjoern~Penning}
\email[]{bjoern.penning@cern.ch}
\affiliation{High Energy Physics Group, Blackett Laboratory, Imperial College, Prince Consort Road, London, SW7 2AZ, United Kingdom}


\begin{abstract}
The mono-jet search, looking for events involving missing transverse energy~($\MET$) plus one or two jets, is the most prominent collider dark matter search. We show that multi-jet searches, which look for~$\MET$ plus two or more jets, are significantly more sensitive than the mono-jet search for pseudoscalar- and scalar-mediated interactions. We demonstrate this in the context of a simplified model with a pseudoscalar interaction that explains the excess in GeV energy gamma rays observed by the Fermi Large Area Telescope. We show that multi-jet searches already constrain a pseudoscalar interpretation of the excess in much of the parameter space where the mass of the mediator~$\mA$ is more than twice the dark matter mass~$\mDM$. With the forthcoming run of the LHC at higher energies, the remaining regions of the parameter space where $\mA>2\mDM$ will be fully explored. Furthermore, we highlight the importance of complementing the mono-jet final state with multi-jet final states to maximise the sensitivity of the search for the production of dark matter at colliders.
\end{abstract}
 
\maketitle

\section{Introduction}
Weakly interacting massive particles (WIMPs) are the most studied and arguably the best motivated candidate for particle dark matter (DM) as they are present in many extensions of the Standard Model (SM). A particularly appealing feature of WIMPs is that they should be detectable with current or near-term experiments~\cite{Bertone:2010at}. 

The plethora of DM models poses a challenge of how to interpret DM searches in a generic way. One approach is to classify the DM model by the particle mediating the interaction. A particularly interesting class of models involves the exchange of a spin-0 $s$-channel scalar or pseudoscalar mediator, since additional scalars and pseudoscalars are a generic prediction of extensions of the Standard Model (SM) Higgs sector. Pseudoscalars are also particularly interesting as they are a common feature in many of the models proposed to explain the spatially extended gamma-ray excess around the Galactic Centre observed with the Fermi Large Area Telescope (Fermi-LAT)~\cite{Goodenough:2009gk, *Hooper:2010mq, *Hooper:2011ti, *Abazajian:2012pn, *Hooper:2013rwa, *Gordon:2013vta, *Macias:2013vya, *Daylan:2014rsa, *Zhou:2014lva, *simonaTalk, *Agrawal:2014oha, Calore:2014xka, *Calore:2014nla}.

While scalars and pseudoscalars with a mass below 10~GeV can be probed by flavour-changing observables at colliders~\cite{Schmidt-Hoberg:2013hba,*Dolan:2014ska}, heavier pseudoscalars whose dominant interaction is with DM are particularly difficult to detect. Pseudoscalar-mediated interactions result in a suppressed tree-level spin-dependent interaction and an unobservably small loop-level spin-independent interaction at direct detection experiments, making this interaction inaccessible for these experiments~\cite{Freytsis:2010ne,*Dienes:2013xya,Boehm:2014hva}. The most prominent collider search for DM production at the Large Hadron Collider (LHC) is the mono-jet search~\cite{Cao:2009uw, *Beltran:2010ww, *Goodman:2010yf, *Goodman:2010ku, *Rajaraman:2011wf, *Fox:2011pm,Bai:2010hh,Khachatryan:2014rra,Aad:2015zva}, which searches for events with a high momentum jet from initial state radiation in combination with significant missing transverse energy~(\met).  As we will demonstrate (see also~\cite{Buckley:2014fba, Harris:2014hga,Haisch:2015ioa}), the mono-jet search has limited sensitivity to pseudoscalar- and \mbox{scalar-mediated interactions.}

In contrast, we show that multi-jet plus $\met$ collider searches significantly extend the sensitivity of the LHC to these interactions. These searches are designed to be inclusive and probe a large region of the topological and kinematic phase space, probing jet-multiplicities~$\ge 2$ with several kinematic variables, including $\met$ and the scalar sum of the jets $p_T$ ($H_T$). Typically, the multi-jet plus $\met$ final state has been used to search for supersymmetry (SUSY) at the LHC. In this letter we demonstrate that this final state also has excellent sensitivity to the pair-production of DM from pseudoscalar and scalar mediators. This is because the production of pseudoscalar- or scalar-mediators is typically dominated by gluon fusion~\cite{Haisch:2012kf,*Fox:2012ru}, which in turn generally leads to events with higher jet-multiplicity in the final state~\cite{Haisch:2013ata}. 

As an example of the utility of the multi-jet plus $\met$ searches, we apply our limits to a pseudoscalar model that can account for the Fermi-LAT excess. We show that current limits already exclude much of the Fermi-LAT excess parameter space where the pseudoscalar's mass~$(\mA)$ is more than twice the DM mass~$(\mDM)$. Future limits at $\sqrt{s}=$13~TeV will fully probe the remaining parameter space where $\mA>2\mDM$.

\section{\label{sec:multi-jet} Mono-jet and Multi-jet searches}

The benchmark pseudoscalar model that we consider is a simplified model following the ansatz of the Minimal Simplified Dark Matter (MSDM) models~\cite{Buchmueller:2014yoa,*Malik:2014ggr}, which have four free parameters: $\mDM$, $\mA$ and two couplings $\gDM$ and $\gSM$. The interaction terms are
\begin{equation}
\label{eq:Lint}
\mathcal{L}_{\mathrm{int}}=i \gDM A \bar{\chi}\gamma^5 \chi+i \gSM \sum_q \frac{m_q}{v} A\, \bar{q}\gamma^5 q\;,
\end{equation}
where $\chi$ is a Dirac fermion, the sum is over all quarks, $m_q$ is the quark mass and $v=246$~GeV is the Higgs vacuum expectation value. Motivated by the Minimum Flavour Violation hypothesis~\cite{D'Ambrosio:2002ex}, we assume that the couplings of the pseudoscalar to quarks are proportional to~$m_q$. With this coupling structure, the $s$-channel production of~$A$ is dominated by gluon fusion. We adopt a simplified model approach in this work as it provides a more accurate framework to characterise the results of collider DM searches when the mediator is light enough to be produced~\cite{Bai:2010hh,Fox:2011pm,Goodman:2011jq,*Shoemaker:2011vi, *Buchmueller:2013dya, *Busoni:2013lha, *Busoni:2014sya, *Busoni:2014haa}. This simple model can explain the Fermi-LAT excess while remaining consistent with other constraints~\cite{Boehm:2014hva}, and is a useful proxy for the structure found in two-Higgs-doublet models (2HDM) and in extended 2HDM that have mixing with a singlet-like pseudoscalar (\`{a} la the NMSSM)~\cite{Gunion:1989we}.

The MSDM ansatz assumes that the pseudoscalar width $\Gamma_A$ is fully determined by the decays to quarks, gluons and DM. The only free parameters affecting $\Gamma_A$ are the four basic parameters $\{\mDM,\mA,\gDM,\gSM\}$. Expressions for $\Gamma_A$ are given in Ref.~\cite{Haisch:2015ioa}. $\Gamma_A$ is dominated by DM and top-quark decay (when kinematically allowed) as $m_q/v$ does not suppress these decays. We will focus on the regime where $\mA>2 \mDM$ in which case $\mathrm{BR}(A\to \chi \bar{\chi})$ is generally large~\footnote{The regime $\mathrm{BR}(A\to \chi \bar{\chi})\ll1$ may be probed in $b\bar{b}+A$ searches, where $A\to \tau^+ \tau^-,\mu^+\mu^-$~\cite{Kozaczuk:2015bea}}. The collider signatures therefore involve the production of $A$ that decays to a pair of DM particles. 

The mono-jet search at the LHC has been advocated as the primary model-independent search for DM production at colliders~\cite{Cao:2009uw, Beltran:2010ww, Goodman:2010yf, Bai:2010hh, Goodman:2010ku, Rajaraman:2011wf, Fox:2011pm}. Events with a second jet may also be allowed but events with three or more are rejected to avoid background contamination from processes with high jet-multiplicity, like top-quark production. 

In contrast, the more inclusive $\AT$~\cite{Chatrchyan:2013lya,*Chatrchyan:2012wa,*Chatrchyan:2011zy,*Khachatryan:2011tk}, $\MT$~\cite{Khachatryan:2015vra,*Chatrchyan:2012jx}, \Razor~\cite{Chatrchyan:2012uea}, or $MHT$-$H_T$~\cite{Collaboration:2011ida,*Chatrchyan:2012lia,*Chatrchyan:2014lfa} multi-jet plus~$\met$ searches place fewer constraints on the phase space. Each event is characterised by the number of jets and hadronic activity $H_T$, as well as other kinematic variables. These bins are combined in a likelihood fit and allow the multi-jet searches to take advantage of different signal-to-background compositions in these numerous search regions to attain better sensitivity. For instance, whilst the CMS mono-jet analysis employs a single inclusive~$\met$ bin with the~$\met$ threshold ranging between 250~GeV and 550~GeV, the $\MT$ search combines more than~$100$ exclusive search regions. Similar kinematic selections and jet-multiplicity categorisations are utilised by the $\AT$ and \textsc{Razor} searches. These inclusive searches are an important pillar of the search strategy for new physics at the LHC, providing the best possible sensitivity to a large variety of SUSY production and decay topologies~\cite{Agashe:2014kda}. So far they have largely been ignored for searches involving the pair production of DM. A \textsc{Razor} search was previously investigated in~\cite{Fox:2012ee,CMS:2015fla} but found a small improvement over the mono-jet search. The search in~\cite{Fox:2012ee} considered one inclusive signal region, which was optimised for models with a (axial-)vector mediator. Most of the events arising from gluon-induced models, like scalar or pseudoscalar exchanges, were rejected. In contrast, our analysis includes all of the accessible signal regions of the~$\MT$ search and maintains excellent performance for a variety of signal models. Unfortunately, Ref.~\cite{CMS:2015fla} did not consider gluon-induced models so no direct comparison is possible.

To determine the sensitivity of these collider searches for our model, we reinterpret the CMS mono-jet~\cite{Khachatryan:2014rra} and $\MT$~\cite{Khachatryan:2015vra} analyses using the \textsc{Powheg Box V2} generator~\cite{Haisch:2013ata,Nason:2004rx, *Frixione:2007vw, *Alioli:2010xd}. This generates, at leading order with exact top-quark mass effects, DM pair production together with one parton via an $s$-channel pseudoscalar mediator.  We use the fixed width approximation, having checked that our results match when the running width is used. We use the \textsc{mstw2008lo} parton distribution functions with renormalisation ($\mu_R$) and factorisation scale ($\mu_F$) set to $\mu/2$, where $\mu = \sqrt{m_{\bar{\chi}{\chi}}^{2} + p_{T,j1}^{2}} + p_{T,j1}$, $m_{\bar{\chi}{\chi}}$ is the invariant mass of the DM pair and $p_{T,j1}$ is the transverse momentum of the leading jet. 
Scale uncertainties on the cross-section were found to be~$\mathcal{O}(\pm 40\%)$~\cite{Haisch:2015ioa}. To be conservative, we do not apply a $K$-factor of $1.6$ as used in~\cite{Buckley:2014fba} to account for higher order corrections, since a computation of the next-to-leading-order corrections with a finite top-quark mass is not available. The events generated by \textsc{Powheg} are interfaced with \textsc{Pythia~8.180}~\cite{Sjostrand:2007gs, *Sjostrand:2006za} for parton-shower effects and hadronisation. Finally, signal events are passed through \textsc{Delphes~v3.2.0}~\cite{deFavereau:2013fsa, *Ovyn:2009tx} for detector simulation.

   \begin{figure}[t!]
\includegraphics[width=0.9\columnwidth]{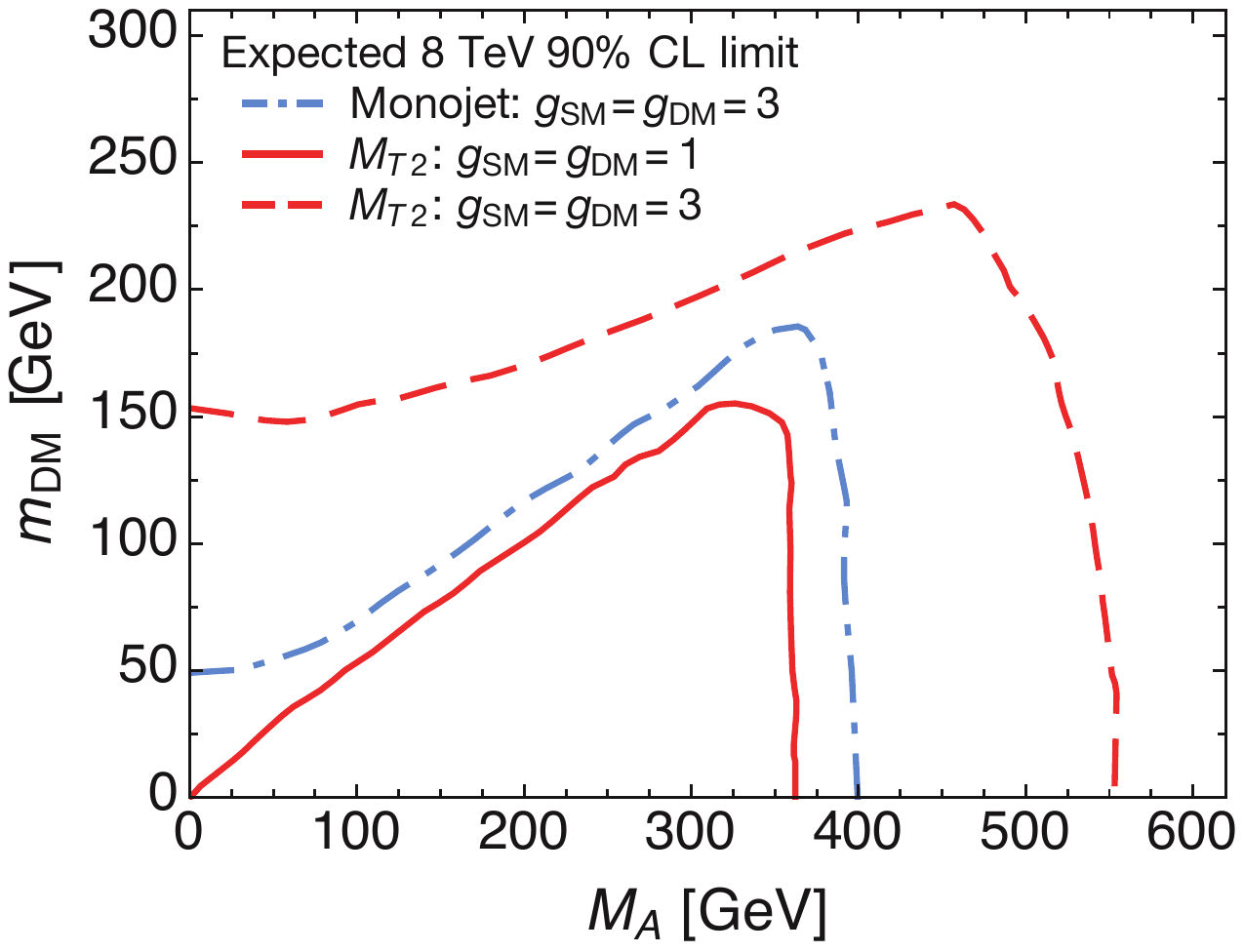}\\
\includegraphics[width=0.9\columnwidth]{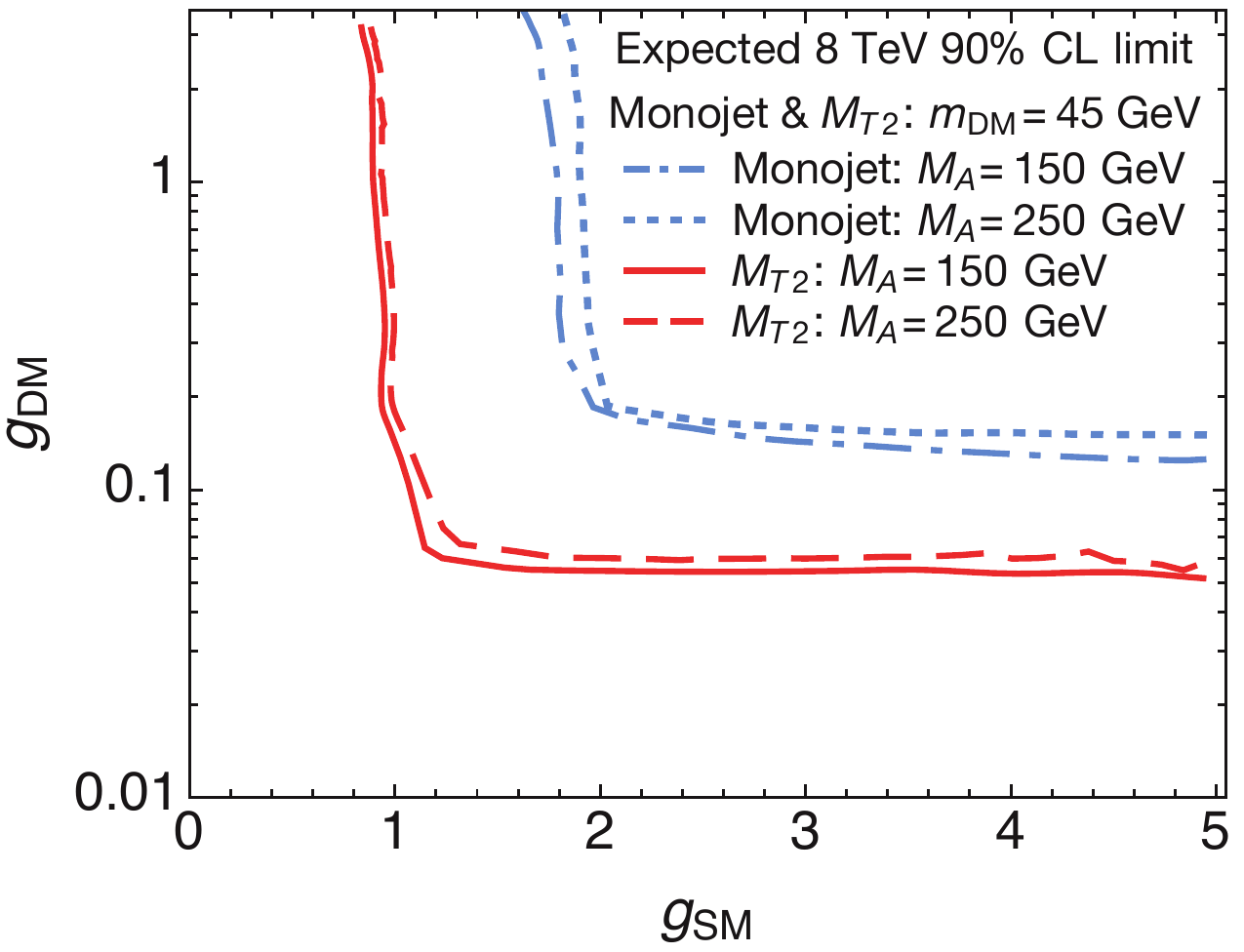}
 \caption{Comparison of the expected $90\%~\mathrm{CL}$ exclusion contours from our mono-jet (blue line) and $\MT$ (red line) analyses. Regions below and above the lines are excluded in the upper and lower panels, respectively. The expected~$\MT$ limits are significantly better than the expected mono-jet limits over the entire parameter space.}
 \label{fig:MA-MDM}
 \end{figure}

Figure~\ref{fig:MA-MDM} shows a comparison of the expected~90\% confidence level~(CL) exclusion contours from our mono-jet and $\MT$ analyses in the $\mA - \mDM$ plane (upper) for two different coupling scenarios, $\gSM=\gDM=1$ and~$3$, as well as in the $\gSM-\gDM$ plane (lower) for $\mDM=45$~GeV and $\mA=150$~GeV and~$250$~GeV. This comparison assumes 20~fb$^{-1}$ at 8~TeV. Both planes show that the~$\MT$ search possesses significantly better expected sensitivity over the mono-jet search. In particular, the~$\MT$ search has the ability to exclude large regions in the $\mA - \mDM$ plane for both $\gSM=\gDM=1$ and~$3$, up to $\mA=350~\mathrm{GeV}$ and~550~GeV respectively for small~$\mDM$. In contrast, for $\gSM=\gDM=1$ the mono-jet analysis does not find any limit while the reach in~$\mA$ for $\gSM=\gDM=3$ is more than 150~GeV less than for~$\MT$. The lower panel shows that for~$\mA>2 \mDM$, the multi-jet plus $\met$ analysis is expected to probe couplings down to $\gDM\simeq0.05$ and $\gSM \simeq 1$. In comparison, the mono-jet search possesses sensitivity for $\gDM\gtrsim 0.12$ and $\gSM \gtrsim 2$, consistent with the findings in~\cite{Haisch:2015ioa}.
 
Based on these results, we conclude that multi-jet plus~$\met$ searches exhibit better sensitivity than the mono-jet analysis over the entire parameter space. The improved sensitivity of~$\MT$ is a result of categorising the search in bins of jet-multiplicity and several kinematic variables, allowing for differences in signal and background in these various categories to be exploited. We find that the low and medium $H_T$ categories for two~jets provide a large fraction of the sensitivity for our pseudoscalar model. However, significant additional sensitivity is gained by the inclusion of low and medium $H_T$ categories with 3-5 jets. The higher-jet bins are particularly important for our model since $\sim60\%$ and $\sim30\%$ of the events in the low and medium $H_T$ categories have gluon-fusion $(gg)$ and quark-gluon $(qg)$ production, which typically produce more jets in the final state. The remaining $\sim10\%$ of events are from quark anti-quark ($q \bar{q}$) or gluon heavy-quark initial states.  Having demonstrated the enhanced sensitivity of~$\MT$ over the mono-jet search, in the following we will show only limits from the~$\MT$ analysis.
 
\section{\label{sec:results} Constraining the Fermi-LAT excess}

\begin{figure*}[t!]
\includegraphics[width=0.67\columnwidth]{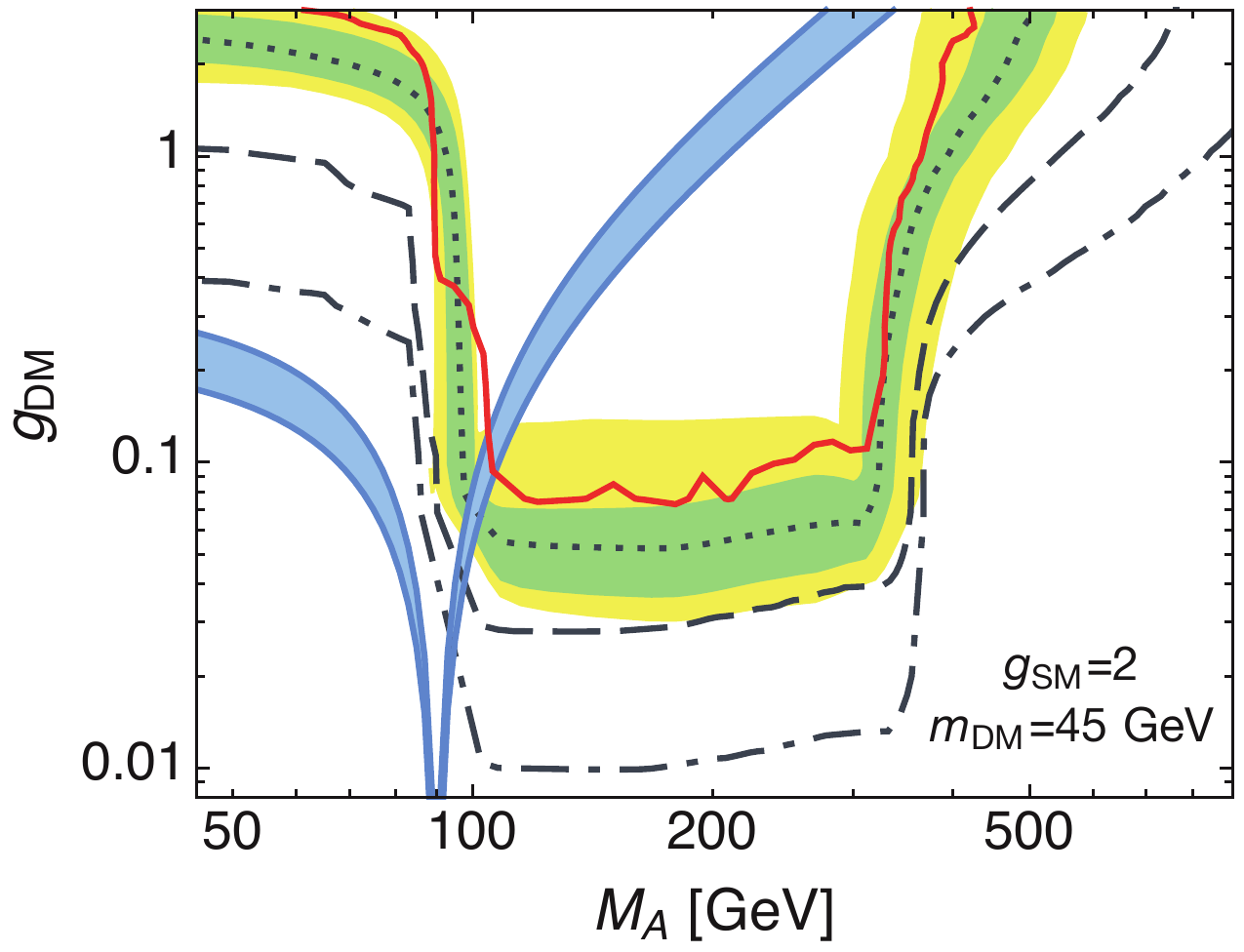}
\includegraphics[width=0.67\columnwidth]{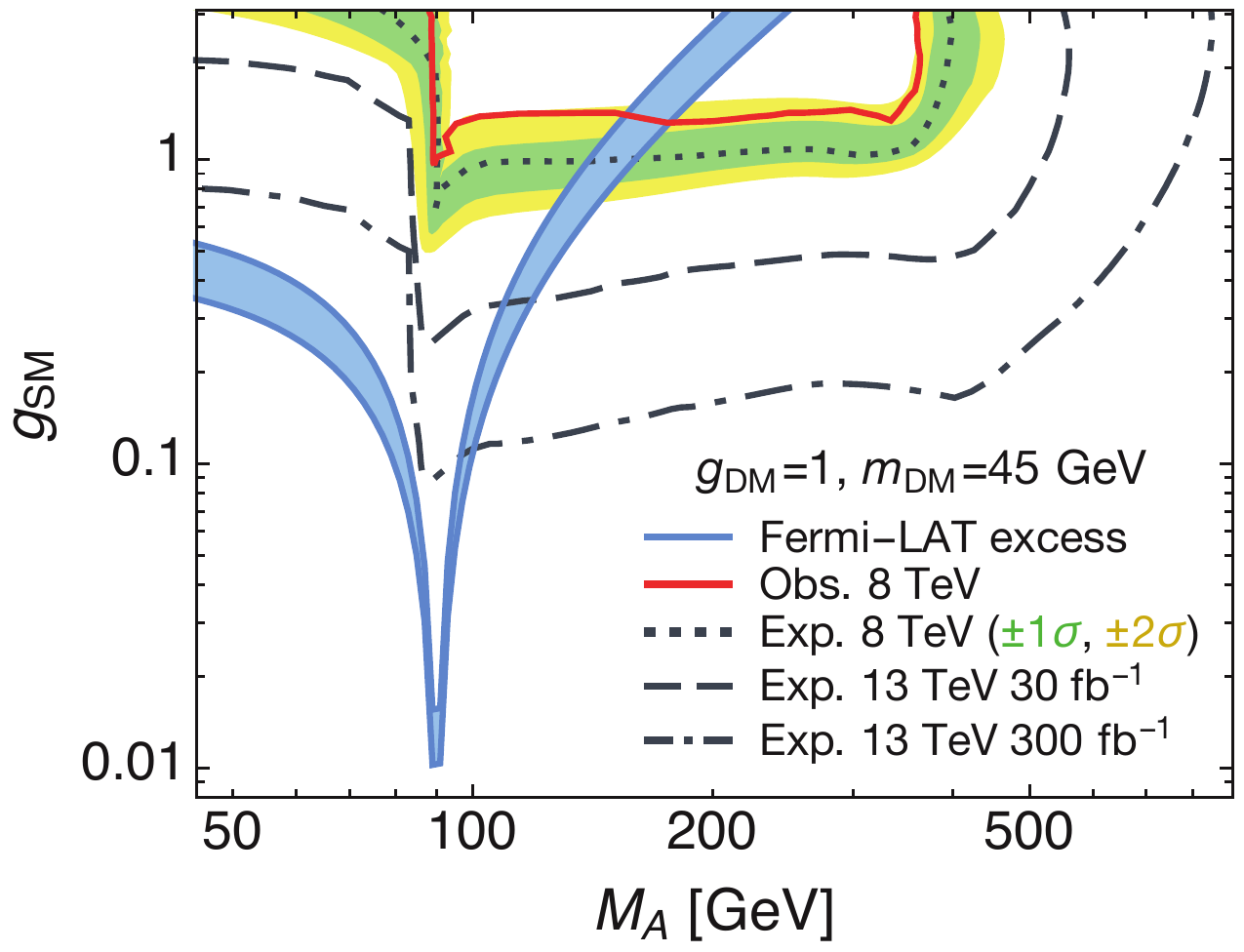}
\includegraphics[width=0.67\columnwidth]{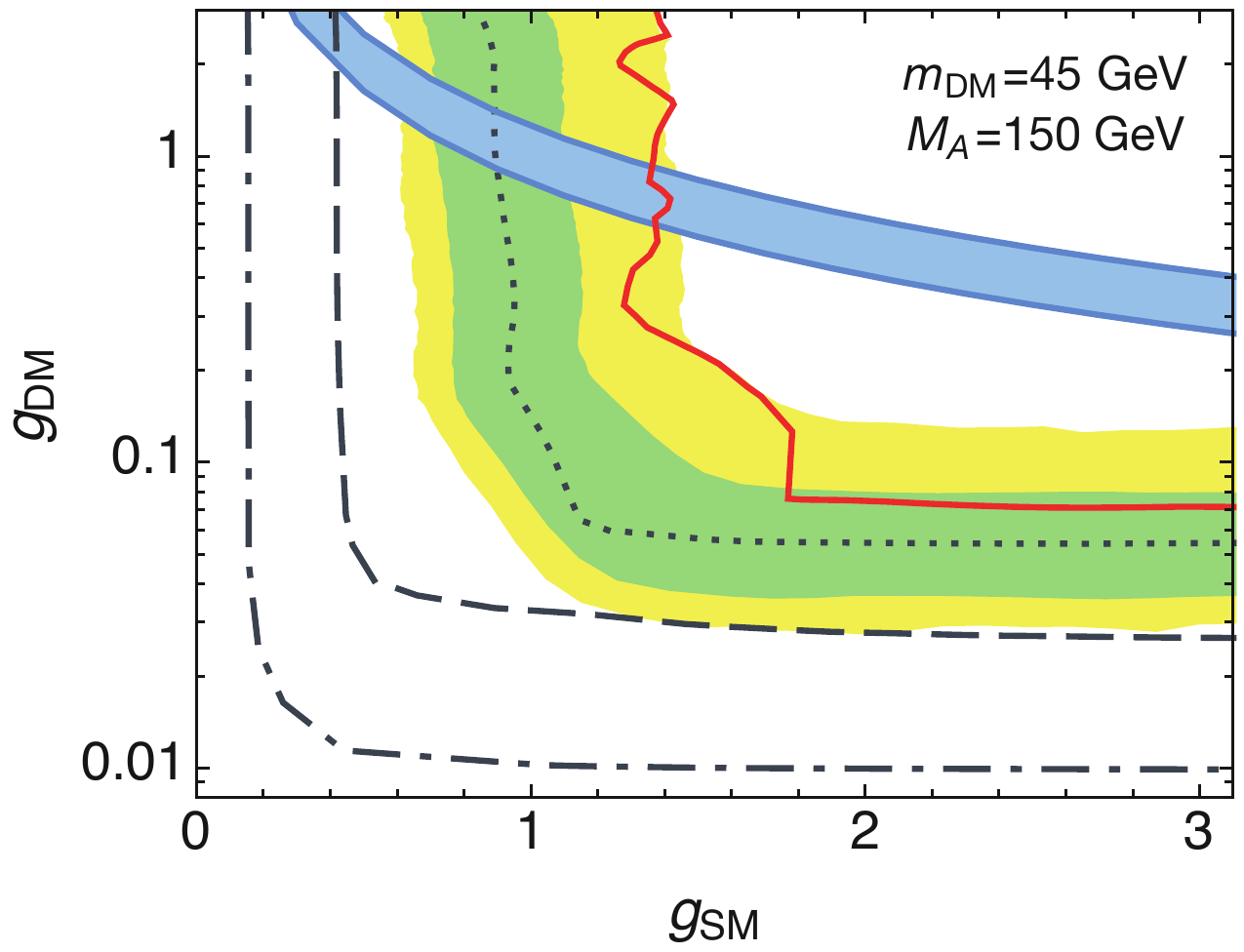}
 \caption{Comparison of the region consistent with the Fermi-LAT excess (blue shaded) and the observed (red) and expected (black dotted) $90\%~\mathrm{CL}$ exclusion contours of the~$\MT$ analysis. The green and yellow shaded regions show the~$\pm1\sigma$ and~$\pm2 \sigma$ bands of the expected~$\MT$ limits. The black dashed and dot-dashed lines show projected limits at~$13~\mathrm{TeV}$. The region above the lines is excluded. In each panel, we fix~$\mDM=45~\mathrm{GeV}$ and one of $\{\mA,\gDM,\gSM\}$, as indicated.\label{fig:results}  }
 \end{figure*}

The Fermi-LAT observation of a spatially extended gamma-ray excess around the Galactic Centre has generated great interest since it may be explained by annihilating DM. Unfortunately, many indirect detection signals, including the Fermi-LAT excess, do not give irrefutable evidence for DM because of large astrophysical uncertainties~\cite{Siegal-Gaskins:2013tga,*Abazajian:2010zy,*Hooper:2013nhl,*Carlson:2014cwa,*Petrovic:2014uda,*Petrovic:2014xra,*Cholis:2015dea}. For instance, Ref.~\cite{Bartels:2015aea,*Lee:2015fea} suggests that the excess could be explained by point sources (PS) that lie just below the current Fermi-LAT threshold. While detecting members of the PS population would corroborate an astrophysical origin for the excess, a complementary signal in direct detection or collider experiments is required to corroborate a DM origin.
  
A plethora of models involving a pseudoscalar mediator have been proposed to explain the Fermi-LAT excess~\cite{Boehm:2014hva,Alves:2014yha,*Hektor:2014kga,*Berlin:2014tja, *Izaguirre:2014vva, *Cerdeno:2014cda, *Ipek:2014gua, *Han:2014nba, *Arina:2014yna, *Cheung:2014lqa, *Huang:2014cla, *Ghorbani:2014qpa, *Cahill-Rowley:2014ora, *Guo:2014gra, *Cao:2014efa, *Freytsis:2014sua, *Cerdeno:2015ega,*Berlin:2015wwa}. As pseudoscalar-mediated interactions are suppressed at direct detection experiments, colliders are the most promising way to independently test a pseudoscalar-mediated explanation for the gamma-ray excess. We therefore investigate the implications of the $\MT$ limits on the model defined by Eq.~\eqref{eq:Lint}, which can explain the Fermi-LAT excess.
 
We fit to the Fermi-LAT excess energy spectrum in~\cite{Calore:2014xka, *Calore:2014nla}, assuming the DM halo follows a generalised NFW profile with $\gamma=1.26$, $r_s=20~\mathrm{kpc}$, $r_{\odot}=8.5~\mathrm{kpc}$ and $\rho_{\odot}=0.4~\mathrm{GeV}\,\mathrm{cm}^{-3}$. We shower the annihilation products with \textsc{Pythia~8.186}~\cite{Sjostrand:2007gs}. For this model, we obtain $\mDM=44.9^{+5.3}_{-4.6}~\mathrm{GeV}$. As in~\cite{Calore:2014xka, *Calore:2014nla}, we find that values up to $\mDM\simeq65$~GeV provide a reasonable fit ($\text{p-value}>0.05$). For~$\mDM=45~\mathrm{GeV}$ and the halo parameters mentioned, the preferred annihilation cross-section is $\sigv=3.2\pm0.4\times10^{-26}~\mathrm{cm}^3\,\mathrm{s}^{-1}$. This is a factor of two larger than values in Ref.~\cite{Calore:2014xka, *Calore:2014nla} because we assume~$\chi$ is a Dirac fermion while Ref.~\cite{Calore:2014xka, *Calore:2014nla} assumed a Majorana fermion.

The annihilation cross-section for $\chi \bar{\chi}\to A\to q \bar{q}$ is
\begin{equation}
\langle \sigma v\rangle_q =\frac{3 m_q^2}{2 \pi v^2}\frac{\gDM^2 \gSM^2 \mDM^2}{(\mA^2-4 \mDM^2)^2+\mA^2 \GamA^2} \sqrt{1-\frac{m_q^2}{\mDM^2}}\;.
\end{equation}
This equation allows us to map $\sigv=\sum_q\langle \sigma v\rangle_q$ to the parameters in our model. The shaded blue bands in Fig.~\ref{fig:results} show the region consistent with the Fermi-LAT excess. In all panels, we assumed~$\mDM=45~\mathrm{GeV}$ and $\sigv=1.4~\mathrm{to}~3.3 \times10^{-26}~\mathrm{cm}^3\,\mathrm{s}^{-1}$. The lower value follows from variations in the halo parameters, principally~$\rho_{\odot}$ which may be as large as~$0.56~\mathrm{GeV}\,\mathrm{cm}^{-3}$~\cite{Read:2014qva} (the annihilation flux~$\Phi$ scales as $\Phi\propto\rho^2_{\rm{DM}}\sigv$). The upper value follows from the Fermi-LAT $95\%~\mathrm{CL}$ upper limit on $\sigv$ from dwarf spheroidal galaxies~\cite{Ackermann:2015zua}. 

To compare the region consistent with the Fermi-LAT excess with the~$\MT$ search, we establish both expected and observed 90\%~CL limits. These are given by the dotted black and solid red lines, respectively in Fig.~\ref{fig:results}. To quantify the compatibility of the expected and observed limits we also determine the expected $\pm1\sigma$ and $\pm2 \sigma$ bands (shaded green and yellow respectively) with a toy experiment technique using the reported background uncertainties in Ref.~\cite{Khachatryan:2015vra}. The expected and observed limits also include a 20\% systematic uncertainty on the signal yield, which is typical for these searches~\cite{Khachatryan:2014rra,Khachatryan:2015vra}. For both bands we have validated our implementation with the~$\MT$ public results and find good agreement. Based on the expected sensitivity of the~$\MT$ search shown in Fig.~\ref{fig:MA-MDM}, we have chosen $\gSM=2$ (left),  $\gDM=1$ (middle) and $\mA=150$~GeV (right) to illustrate the constraints the~$\MT$ search places on the Fermi-LAT excess. A resonance feature when $\mA\approx 2\mDM$ is seen in both the Fermi-LAT region and the $\MT$ limit in the $\mA - \gDM $ and $\mA - \gSM$ planes. Outside this region, the excess is consistent with $\gDM\sim\gSM\sim\mathcal{O}(1)$.  Owing to the off-shell suppression of the production cross-section, these searches cannot place relevant constraints on the region below~$\mA<2\mDM$.
 
Our observed limit for the~$\MT$ search is approximately~$2\sigma$ weaker than our expected limit. This is compatible with~\cite{Abdallah:2014hon}, where the observed limit for direct production of light squarks is also weaker than expected.  In contrast, expected and observed limits are similar for the mono-jet analysis. This suggests that the weaker limit is caused by statistical fluctuations in the background estimates in some of the phase space regions probed by the~$\MT$ search that are inaccessible to the mono-jet search. Of course, we cannot exclude the possibility that a DM signal causes the weaker limit, but given that this is a $2\sigma$ effect, additional data are required to draw any significant conclusion. 

Even with a $\sim2\sigma$ weaker observed limit than expected, the~$\MT$ search still excludes a significant fraction of the Fermi-LAT excess region for $2 \mDM \lesssim \mA \lesssim 400~\mathrm{GeV}$. For $\gSM=2$, $\MT$ excludes all of the excess region above $\mA =107~\mathrm{GeV}$ (left panel), while for $\gDM=1$ mediator masses compatible with the excess above $177~\mathrm{GeV}$ (middle panel) are excluded.  The right panel shows that~$\MT$ is able to exclude all of the excess region for $\gDM<0.93$ for an illustrative mediator mass of $\mA=$150~GeV. In these panels we assumed that $\mDM=45$~GeV but similar conclusions are found for values up to $\mDM=65$~GeV. In fact, Fig.~\ref{fig:MA-MDM} demonstrates that the~$\MT$ limits have little dependence on~$\mDM$ for $\mDM\lesssim125$~GeV.
 
To illustrate how the Fermi-LAT excess parameter space might be probed in the future, we also provide projected sensitivities of the~$\MT$ search. The basis for these extrapolations are the 8 TeV limits, which are rescaled assuming that the underlying performance of the search in terms of signal efficiency and background suppression remains unchanged. These assumptions were also used in Ref.~\cite{Buchmueller:2014yoa,Malik:2014ggr} and form the basis of \textsc{Collider Reach}~\cite{colliderreach}. Figure~\ref{fig:results} shows the projected limits for an early start-up scenario assuming 13~TeV and $30~\mathrm{fb}^{-1}$ (black dashed) and a long-term scenario with 13~TeV and $300~\mathrm{fb}^{-1}$ (black dot-dashed).  The increase in energy and luminosity will enable this search to significantly increase its sensitivity. Assuming that search performance is maintained, it will be possible to probe almost all of the region $\mA>2\mDM$ compatible with the Fermi-LAT excess.

Finally, we see that the projected limits do not constrain the region~$\mA<2\mDM$. This implies that this search will not be able to probe the `cascade-annihilation' models that explain the Fermi-LAT excess (see e.g.~\cite{Boehm:2014bia,*Ko:2014gha,*Abdullah:2014lla,*Martin:2014sxa,*Berlin:2014pya,*Liu:2014cma,*Gherghetta:2015ysa,*Rajaraman:2015xka}).   In these models, a pair of mediating particles are produced on-shell, requiring $\mA<\mDM$.
 
\section{\label{sec:conclusion} Discussion}


Although the mono-jet search is the most prominent search for DM at the LHC, we have shown that the multi-jet plus~$\MET$ search,~$\MT$, provides more stringent constraints on DM production for a pseudoscalar mediator. The additional sensitivity of the multi-jet search originates from binning the search into categories of jet-multiplicity and kinematic variables like $H_T$ and $\MT$, as well as from extending to higher jet-multiplicities than the one- or two-jet final state probed by the mono-jet search. This is especially relevant for gluon-fusion produced signal models, including the pseudoscalar model discussed here and models with a scalar mediator. \textsc{Powheg} has the exact top-quark mass dependence for DM pair production and one parton in the final state. Topologies with higher jet-multiplicities rely on the parton shower for additional jets, which could introduce uncertainty not fully accounted by our analysis. To estimate the impact of the parton shower producing too many energetic jets, we performed a re-weighting of the jet-multiplicity distribution of the signal events. As an extreme variation, we consider the scenario where all~3-5 jet events are moved to the two-jet categories (while keeping other kinematic variables fixed) and find that the resulting limit is still contained in the expected $\pm1\sigma$ bands in Fig.~\ref{fig:results}. The robustness of the $\MT$ search against such variations arises from the design criteria of inclusive searches, which typically require that final states with differing jet-multiplicity have a similar sensitivity. Therefore, signal events assigned to the wrong category contribute to the analysis with a similar weight and thus maintain the overall performance of the search. Although we find that our conclusions are unchanged by this re-weighting, it would be highly desirable to have theoretical tools that include the full top-quark mass dependence in events with multiple partons in the final state.

The Fermi-LAT gamma-ray excess remains an enigma. It may be straightforwardly explained with simple DM models involving the exchange of an $s$-channel pseudoscalar mediator, but unfortunately, the Fermi-LAT data are not sufficient to exclude mundane explanations with astrophysical sources. A DM signal in a complementary experiment is required to confirm a DM origin. Confronting the~$\MT$ limits against parameter space favoured by the Fermi-LAT excess shows that the LHC provides crucial input on pseudoscalar models. We demonstrated that for $\mA > 2 \mDM$, much of the parameter space is already constrained. Our 13~TeV projections indicate that essentially all of the region $\mA > 2 \mDM$ will be probed by the next LHC run. 

As multi-jet searches consider events with at least two jets in the final state, the overlap with mono-jet searches, which allow up to two jets in the final state, is small. For this reason, we strongly recommend that the phase space covered by mono-jet and multi-jet searches is combined in a single search to further improve the sensitivity of the LHC to DM production.

\begin{acknowledgments}
\paragraph{Acknowledgments.} The authors are grateful for discussions at the IPPP Senior Fellowship Meeting at the University of Bristol, and to Matt Dolan, Uli Haisch and Tim Tait for comments on an early version of this letter. The work of O.B.\ and S.M.\ is supported in part by the London Centre for Terauniverse Studies (LCTS), using funding from the European
Research Council via the Advanced Investigator Grant
267352. The work of B.P.\ is supported by an Imperial College Junior Research Fellowship. C.M.\ is supported by the ERC starting grant {\it WIMPs Kairos} and is grateful to MITP for its hospitality and partial support during parts of this work.
\end{acknowledgments}

\bibliographystyle{apsrev4-1}
\bibliography{ref}

\end{document}